\documentclass[aip,onecolumn, superscriptaddress,showpacs,floatfix]{revtex4-1}
\newcommand{\noop}[1]{}
\usepackage{dcolumn}
\usepackage{amssymb}
\usepackage{amsmath}
\usepackage{graphicx}
\usepackage{placeins}
\usepackage{bm}
\usepackage[T1]{fontenc}
\usepackage{color}
\usepackage{url}
\usepackage{rotating}
\usepackage{todonotes}
\usepackage[squaren]{SIunits}
\usepackage{wasysym}
\usepackage[latin1]{inputenc}
\usepackage[T1]{fontenc}
\usepackage{ae,aecompl}

\usepackage{xcolor}
\newcommand{\nn}{\nonumber}

\begin{document}

\lineskip .2em

\title{The impact of model detail on power grid resilience measures}

\author{Sabine Auer}
\email{auer@pik-potsdam.de}
\affiliation{Potsdam Institute for Climate Impact Research, 14412 Potsdam, Germany}
\affiliation{Department of Physics, Humboldt University Berlin, 12489 Berlin, Germany}
\author{Kirsten Kleis}
\affiliation{Oldenburg University, Germany}
\author{Paul Schultz}
\affiliation{Potsdam Institute for Climate Impact Research, 14412 Potsdam, Germany}
\affiliation{Department of Physics, Humboldt University Berlin, 12489 Berlin, Germany}
\author{J\"urgen Kurths}
\affiliation{Potsdam Institute for Climate Impact Research, 14412 Potsdam, Germany}
\affiliation{Department of Physics, Humboldt University Berlin, 12489 Berlin, Germany}
\affiliation{Institute of Complex Systems and Mathematical Biology, University of Aberdeen, Aberdeen AB24 3FX, UK}
\affiliation{Department of Control Theory, Nizhny Novgorod State University, 606950 Nizhny Novgorod, Russia}
\author{Frank Hellmann}
\affiliation{Potsdam Institute for Climate Impact Research, 14412 Potsdam, Germany}

\date{\today}

\begin{abstract}
Extreme events represent a challenge to natural as well as man-made systems. For critical infrastructure like power grids, we need to understand their resilience against large disturbances. Recently, new measures of the resilience of dynamical systems have been developed in the complex system literature. Basin stability and survivability respectively assess the asymptotic and transient behavior of a system when subjected to arbitrary, localized but large perturbations.
To employ these methods to assess the resilience of power grids, we need to choose a model of the power grid. So far the most popular model that has been studied is the classical swing equation model for the frequency response of generators and motors. In this paper we study a more sophisticated model of synchronous machines that also takes voltage dynamics into account, and compare it to the previously studied model. This model has been found to give an accurate picture of the long term evolution of synchronous machines in the engineering literature for post fault studies.
We find evidence that some stable fix points of the swing equation become unstable when we add voltage dynamics. If this occurs the asymptotic behavior of the system can be dramatically altered, and basin stability estimates obtained with the swing equation can be dramatically wrong. We also find that the survivability does not change significantly when taking the voltage dynamics into account. Further, the limit cycle type asymptotic behaviour is strongly correlated with transient voltages that violate typical operational voltage bounds. Thus, transient voltage bounds are dominated by transient frequency bounds and play no large role for realistic parameters.
\end{abstract}

\maketitle

\section{Introduction}
In this paper we study the question of how detailed a dynamical model of the power grid needs to be to accurately assess the impact of extreme events. Power grids are among the most critical infrastructure for modern societies, in particular, power grid failures have dramatic economic and societal impacts. They can shut down transportation and communication networks, force hospitals to work on backup power, and generally bring a modern society to a complete stand still.

As a result, the stability and resilience of power grids is a well studied issue. It usually comes in two forms. First, linear stability measures assess the stability of the operating state of the power grid to inevitable and omnipresent small fluctuations. Secondly, detailed fault simulations ensure that any one component of the power grid can fail while the network remains operational, this is the \textit{(N-1)}-criterium.

The concepts of {\it basin stability} and {\it survivability} of dynamic systems offer a third perspective on the inherent resilience of power grids. These assess the ability of the power grid to withstand localized arbitrary large disturbances \cite{Menck2013,2015arXiv150601257H}. In order to evaluate the resilience of power grids of a particular topology against such large disturbances, a dynamic model of the power grid is required. In the engineering literature, a number of models for power grids of various degrees of accuracy have been developed \cite{Sauer2006,Weckesser2013}. A more detailed model that included some voltage dynamics was studied in \cite{Schmietendorf}. So far, it had not been studied which level of model detail is actually required to assess the response of large networks to large generic disturbances. This paper starts to fill this gap by comparing a 4th order model found to be sufficient for the post fault state analysis in \cite{Weckesser2013} to the classic swing equation model that has been the focus of most of the theoretical work so far.

We find that taking the voltage dynamics into account does not lead to a large change in the transient frequency behavior, but may dramatically change the the asymptotic behavior of the equations.

In the next section we will describe two power grid models of different detail or order. The \emph{swing equation} is the model used overwhelmingly in the theoretical physics literature. The \emph{4th-order model} is a more detailed model separating the electric and mechanical aspects of the power grid to some degree. This was found in the engineering literature to give a better picture of the long term dynamics of power generators. We will then briefly review the synthetic power grid topologies we use in this paper.

In the subsequent section we briefly review the methods of basin stability and survivability that we will study. Finally, we present our results before concluding and discussing further steps.

\section{Power grid models}
\subsection{Swing equation}
The swing equation describes the power grid dynamics of $N$ synchronous machines with two equations per node: for phases $\phi_i$ and frequency deviations $\omega_i$.
In this so-called  classical model, generators are represented as constant power, constant voltage sources \cite{Nishikawa2015,Anderson2003} with voltage magnitude $U_i$ and rotating complex voltage $V_i=e^{i\phi_i}U_i$. Besides the constant voltage magnitude the machines are parametrized by the constant mechanical input power $P_{i}$, the moment of inertia $H_i$ and an effective damping term $D_i$. The frequency and phase are the instantaneous speed and position of both the electric field voltage and the rotating mass. Loads are assumed to be constant impedances so that they can be reduced into an effective network structure, where the loads are absorbed into the effective power $P_i$ at a node. This can thus be positive or negative, and the sum of input powers $P_i$ is zero. The admittance matrix of the effective network is called $Y_{ij}$ \cite{Nishikawa2015}, and we set the diagonal such that the row sums are zero, $Y_{ii} = - \sum_j Y_{ij}$. These assumptions allow a fairly accurate description of the system's transient behavior after a disturbance in the time period of the first swing which is usually one second or less \cite{Anderson2003}. The swing equation describes the dynamics of such a deviation, $\omega_i$, from the grid frequency $\omega_{n}$. That is, the instantaneous speed of rotation is $\tilde \omega_i = \omega_i + \omega_{n}$, normal operations are characterized by $\omega_i = 0$. The main content of the system is in equation \eqref{eq:energy_balance}, which is a first order approximation of energy conservation, with power input, the power balance with the rest of the power grid, and a friction term:
\begin{align}
\frac{d\phi_i}{dt} & = \omega_i\;,\\
\frac{2H_i}{\omega_{n}}\frac{d\omega_i}{dt} &= P_i -\Re \sum_{j \neq i} (V_i I^*_{ij}) -D_i\omega_i\;.\label{eq:energy_balance}
\end{align}
where $\Re(V_i I^*_{ij})$ is the real part of the power flow between node $i$ and node $j$.
The complex current $I_{ij}$ from node $i$ to node $j$ is given by:
\begin{equation}
I_{ij} = Y_{ij} (U_i e^{i\phi_i} - U_j e^{i\phi_j})\;.
\end{equation}
It is convenient to introduce the current $I^c_{i} = e^{- i \phi_i} \sum_j I_{ij}$ in the co-rotating frame. The total co-rotating current at a node is given by:

\begin{align}
I^c_i &= e^{- i \phi_i} \sum_{j=1}^{N} Y_{ij} V_j = e^{- i \phi_i} \sum_{j=1}^{N} Y_{ij} U_j e^{i \phi_j} = \sum_{j=1}^{N} Y_{ij} U_j e^{i (\phi_j - \phi_i)} \label{eq:4Ord_current}
\end{align}

As we have $Y_{ii} = - \sum_j Y_{ij}$ we can combine these equations into the swing equation:
\begin{align}
\frac{d\phi_i}{dt} & = \omega_i \label{eq:swingequation_theta}\;,\\
\nonumber \frac{2H_i}{\omega_{n}}\frac{d\omega_i}{dt} &= P_i -\Re(U_i (I^c)^*) - D_i\omega_i \\
 &= P_i -\Re \sum_j U_i Y^*_{ij} U_j e^{i(\phi_i - \phi_j)} - D_i\omega_i \label{eq:swingequation_omega}\;.
\end{align}

The impedance matrix $Y_{ij}=G_{ij}+i B_{ij}$ can often be approximated by $Y_{ij}\approx i B_{ij}$ as the Ohmic resistance of transmission  lines can be disregarded. In this important case, the swing equation reduces to the familiar form:

\begin{align}
\frac{d\phi_i}{dt} & = \omega_i \label{eq:swingequation_theta_final_0}\;,\\
\frac{2H_i}{\omega_{n}}\frac{d\omega_i}{dt} &= P_i - \Re \sum_j U_i B_{ij} U_j (i e^{i(\phi_j - \phi_i)})^*) - D_i\omega_i \label{eq:swingequation_omega_final_1} \\
&= P_i - \sum_j U_i B_{ij} U_j \sin(\phi_i - \phi_j) - D_i\omega_i \label{eq:swingequation_omega_final_2}\;.
\end{align}

The fixed point equations of the dynamics simplify to the power flow equations:

\begin{align}
\omega_i^\star & = 0 \;,\\
P_i &= \sum_j U_i B_{ij} U_j \sin(\phi_i^\star - \phi_j^\star)\;.
\end{align}

\subsection{4th-order model}

Usually the swing equation is used for short time periods to analyze the transient behavior of generators in a power grid, the so-called first swing. The fourth-order model (Eqns.~\ref{eq:4Ordtheta} -- \ref{eq:4OrdEd}) also takes the back reaction of the power flow onto the voltage into account. This has the effect that the voltage as seen by the power grid, and the rotating mass are no longer the same but become dynamically coupled. The voltage is described in a co-rotating frame with axes labeled $d$ and $q$. Thus we have the voltages $E_{q,i}$ and $E_{d,i}$ (see Eqn.~\ref{eq:4OrdEq} and Eqn.~\ref{eq:4OrdEd} respectively), and the complex voltage $U_i = E_{q,i} + i E_{d,i} = e^{-i \phi_i} V_i$ in the co-rotating frame is now dynamical. For convenience we also introduce the notation
\begin{equation}
I^c_i = I_{q,i} + jI_{d,i}
\end{equation}
for the co-rotating current introduced above.

Now the equations for the swing mass are unchanged, being merely energy conservation and phase shift:

\begin{align}
\frac{d\phi_i}{dt} &= \omega_i \label{eq:4Ordtheta}  \\
\frac{2H_i}{\omega_{n}}\frac{d\omega_i}{dt} &= P_i - \Re \sum_{j \neq i} (V_j I^*_{ij})- D_i\omega_i \\
 &= P_i - \Re (U_i (I^c_{i})^*) - D_i\omega_i \label{eq:4Ordomega} \\
 &= P_i - E_{d,i} I_{d,i} - E_{q,i} I_{q,i} - D_i\omega_i \label{eq:4Ordomega2}
\end{align}

but they are now complemented by two equations for the complex voltage,

\begin{align}
T_{d,i} \frac{dE_{q,i}}{dt} &= -E_{q,i} + X_{d,i} I_{d,i}+E_{f,i} \;,\label{eq:4OrdEq} \\
T_{q,i} \frac{dE_{d,i}}{dt} &= -E_{d,i} + X_{q,i} I_{q,i} \;.\label{eq:4OrdEd}
\end{align}

The new parameters have the following physical interpretation: The $E_{f,i}$ are the reference voltage at which the generator is run. The time constants $T_{d,i}$, $T_{q,i}$ parametrize the speed of the voltage dynamics in the d- and q-axis. Finally the reactances $X_{d,i}$, $X_{q,i}$ parametrize the influence of the currents in the generator on the voltage.

The limit towards the swing equation is provided by setting $E_{q,i}=E_{f,i}$ and $E_{d,i}=0$, and ensuring that $\frac{dE_{q,i}}{dt} = \frac{dE_{q,i}}{dt} = 0$. This occurs in the limit 
\begin{equation}
\frac{X_{d/q,i}}{T_{d/q,i}} \rightarrow 0\;.
\end{equation}

For the fixed point of the 4th order model we require more than just the power flow balancing:

\begin{align}
\omega_i^\star &= 0 \;,\\
P_i &= E_{d,i}^\star I_{d,i}(E_d^\star,E_q^\star,\phi^\star) - E_{q,i}^\star I_{q,i}(E_d^\star,E_q^\star,\phi^\star) \;,\\
E_{q,i}^\star  &= X_{d,i} I_{d,i}(E_d^\star,E_q^\star,\phi^\star) +E_{f,i} \;, \\
E_{d,i}^\star &= X_{q,i} I_{q,i}(E_d^\star,E_q^\star,\phi^\star) \;.
\end{align}

In appendix~\ref{sec:app} we provide a derivation of the form of the equation used here, from the form in the engineering literature, which allows us to use reference numerical values for the various parameters introduced.

\subsection{Synthetic power grid topologies}
The before mentioned grid models are run on real but also on artificial power grids in order to  test them on a statistic ensemble of grid topologies. 

In addition to the topology of the Scandinavian power network, we further consider a recently published 
model for synthetic power grid topologies \cite{Schultz2014}. It is a network growth model aiming to
reproduce topological properties of real power grids and other spatially embedded infrastructure networks.
The growth process is controlled by a heuristic redundancy/cost optimization function, which takes not only 
the length of transmission lines but also additionally created redundancy in the form of alternative routes into account.
For a detailed discussion of the algorithm, we refer to \cite{Schultz2014}. Note that the resulting power grid topologies come 
with a spatial embedding of the network and hence information about the link lengths. This allows to 
estimate appropriate admittances $Y$ from textbook parameter values. For comparability with the basin stability literature we choose
to use a constant coupling for this work though.

\section{Stability measures}

In the engineering literature a number of sophisticated stability measures for assessing the stability of post fault states are known. One example of these is the Equal Area Criterion (EAC) that was used to compare different model details in \cite{Weckesser2013}. The EAC allows to assess information about grid stability in real time to prevent system break down. It investigates if the system is capable of absorbing the kinetic energy change induced by a  disturbance in electric power \cite{Pavella2000}. In order to be usable as a real time preventive measure, the EAC avoids full time-domain simulations.

The linear stability of a particular operational state, assumed to be a fix point of the grid model dynamic equations, is given by the largest non-zero eigenvalue of the linearized dynamics around the fix point. A convenient way to study the spectrum of a linear dynamic on a network is the master stability function approach \cite{Nishikawa2015,pecora1998master}. This approach separates out the local dynamics and the network structure.
As the general shape of the master stability function is independent of the actual network, it is possible to quickly evaluate the  asymptotic stability of a given dynamical system for various topologies.
However, problems arise if the Laplacian is not diagonalizable \cite{Wei2011}, e.g. if the ohmic resistances of transmission lines are not neglected.

The first of these methods evaluates the post-fault state and the second the inherent linear stability of the system against small perturbations. In contrast, the measures we will discuss now assess the stability of the system against large, random perturbations at single nodes of the network.

\subsection{Basin stability}
The basin stability (BS) of a multi-stable dynamical system with trajectories $x(t)$ is the fraction of trajectories that approach a desired set of attractors $X^\star$ \cite{Menck2013}. More formally, given a region in phase space $X^0$ that contains our generic perturbations, the basin of attraction within $X^0$ is then $X^{BS} = \{x(0) \in X^0| \lim_{t \rightarrow \infty} x(t) \in X^\star\}$. The basin stability is the ratio of the volumes \begin{equation}\mu_B = \frac{|X^{BS}|}{|X^0|}\; .\end{equation}
In the case of power grids we define our set of desirable attractors to be exactly those that are stationary, that is, $\omega^\star = 0$. The generic perturbations we study depend on the initial operating state of the system $(\phi^\star, \omega^\star)$ or $(\phi^\star, \omega^\star, E^\star)$. They are constructed by taking an arbitrary phase space perturbation $\delta \phi \in [-\pi, \pi]$ and $\delta \omega \in [- \omega_{max}, \omega_{max}]$ and adding them to a single entry (for single node basin stability only one node at a time is perturbed) in the vectors $\phi^\star$ and $\omega^\star$ respectively. That is: $\phi_i(0) = \phi^\star_{i} + \delta_{ij} \delta \phi$, $\omega_i(0) = \omega^\star_{i} + \delta_{ij} \delta \omega$, and $E(0) = E^\star$. We choose not to perturb $E$ in order to facilitate comparisons between the swing equation and the 4th-order model.

\subsection{Survivability}

Survivability measures the ability of a system to keep within some predefined operating regime when experiencing large perturbations. For the power grid, this generally means we want to keep the frequency deviation below $\omega_{crit} = 0.2$Hz before controls kick in. For our purposes we investigate a number of different frequency and voltage thresholds, that are more forgiving than ones used in practice. The surviving region $X^S$ of the system is defined as those initial conditions whose trajectories never violate these bounds. Thus in our case we have: $X^{S} = \{(\phi(0), \omega(0)) \in X^0| \max_{t} |\omega(t)| < \omega_{crit}\}$. We construct the initial conditions through perturbing a single node again, choosing $\omega_{max} = \omega_{crit}$. The survivability is then given by \begin{equation}\mu_S = \frac{|X^{S}|}{|X^0|}\; .\end{equation}

In contrast to basin stability, which does not depend on the transient behavior of the system, the survivability is concerned with the entire trajectory. This can be considered a more realistic measure for power grids where large transient deviations could damage the power grids and require manual intervention to bring it back to an acceptable operating regime.

Given a perturbation that leaves the system within the fixed point's basin of attraction, the maximum frequency deviation is typically given by the first swing. As this is the behavior that is well modeled by the swing equation, we expect that the Swing equation might be a good approximation for survivability.

Finally, for the fourth order model we have the choice to introduce voltage bounds that should not be violated. In real power grids the acceptable voltage thresholds are typically $\pm 0.1$ per unit (pu). In our model even the stable states tend to violate this bound already. We will see later that the voltage bounds are not important relative to the frequency bounds but they do play a large role for the multi-stability of the system though.

\section{Results}

We will begin by showing the general differences in the dynamics of the different grid models by discussing representative time series, the fixed points and the maximum eigenvalues of the system's Jacobian in Section \ref{sec:res_modeldiff}. Afterwards, in Section \ref{sec:res_bs_surv} we compare the impact of model detail on the basin stability and survivability measure for different grid topologies. As the fourth order model increases the model detail by the inclusion of voltage dynamics, in Section \ref{sec:volt_bounds} we particularly focus on voltage bounds in the stability paradigm.

\subsection{General model differences}\label{sec:res_modeldiff}

To perturb a system it is necessary to first identify its fixed points. Starting with a random dispatch scenario, that is, a random distribution of generator ($P_i = +1$) and consumers ($P_i = -1$) across the grid, we try to find a stationary solution for the swing equation (see eq. \eqref{eq:swingequation_omega_final_1}). If there is no solution another random distribution of sinks and sources is chosen.

Having found such a fixed point we use the state with frequency and phase given by the swing equation fixed point and $E_d = 0$ and $E_q = E_f$ as the starting point of a search for a fixed point of the 4th order equation. 

Remarkably, for some dispatch scenarios and networks, this search fails. There are dispatches which according to the swing equation allow for dynamically stable power transport, but are dynamically unstable under the voltage dynamics. This indicates that fixed points that are stable in the swing equation can become unstable when the coupling of the voltage is taken into account.

In order to proceed with the comparison we choose dispatch scenarios that are dynamically stable for both, the swing equation and the 4th order equation. We then considered single node perturbations as described above.

\begin{figure}[tbph!]
\centering
\includegraphics[width=0.49\textwidth]{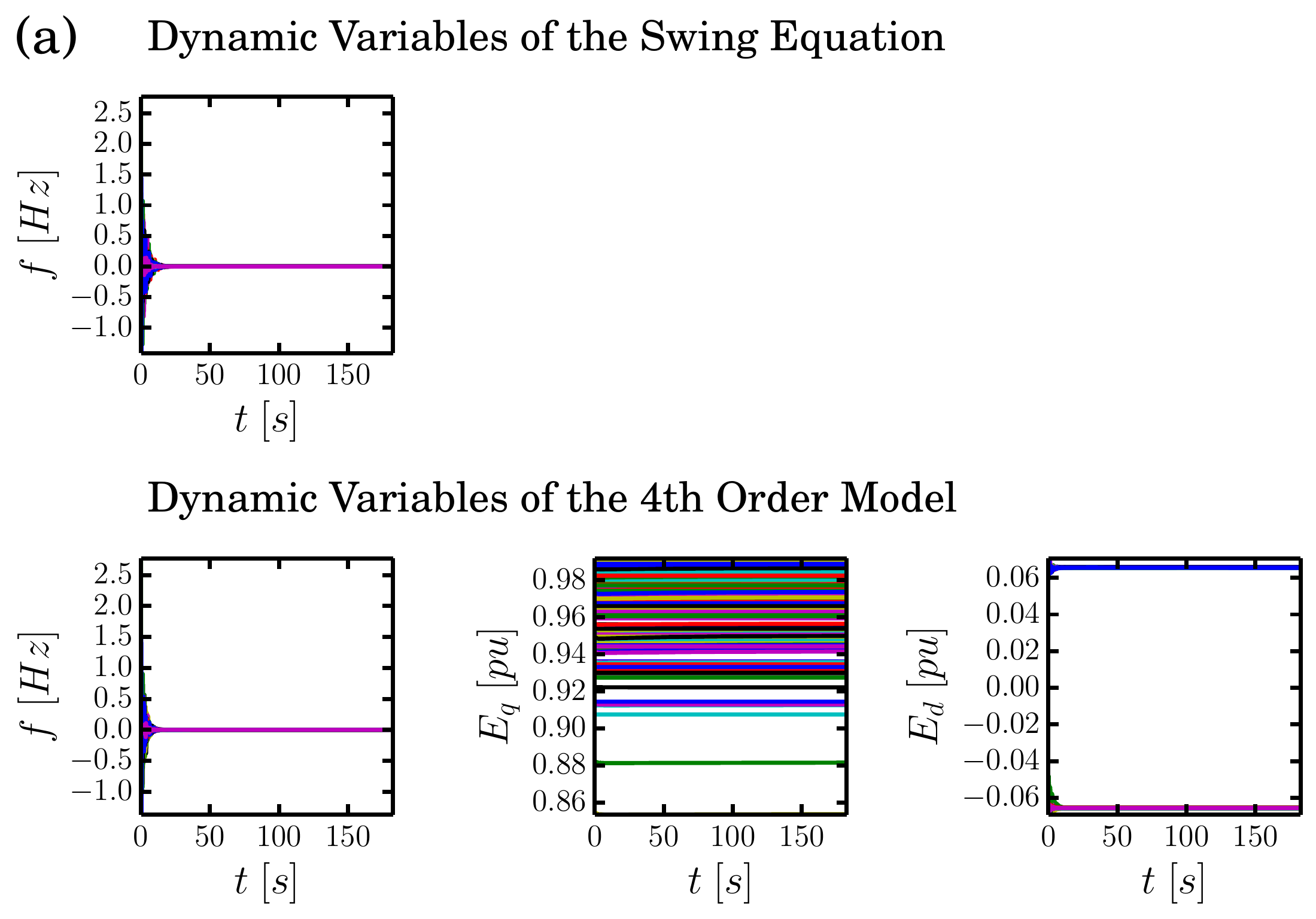}
\includegraphics[width=0.49\textwidth]{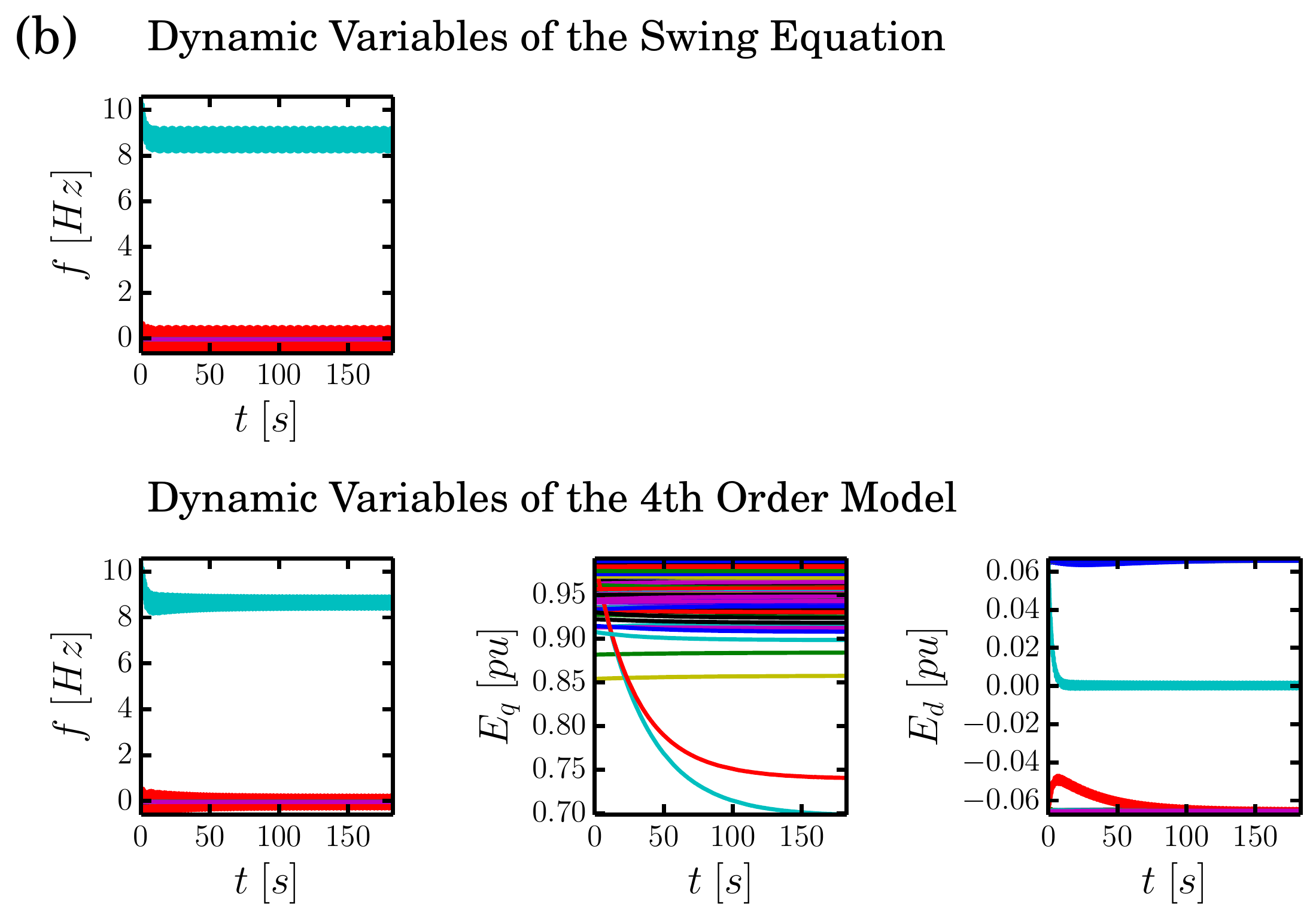}
\caption{Example trajectories for the Scandinavian power grid are (a) fixed point convergence and (b) the limit cycle for random single node perturbations with $\phi_{185}(0)=-0.38$ rad, $f_{185}(0)=-10.9$Hz and $\phi_{199}(0)=-3.00$ rad,    $f_{199}(0)=-10.6$Hz respectively. The colors of the trajectories represent the different grid nodes. Typically, the 4th order follows similar trajectories as the swing equation but converges faster.}
\label{fig:time_plots1}
\end{figure}

\begin{figure}[tbph!]
\includegraphics[width=0.49\textwidth]{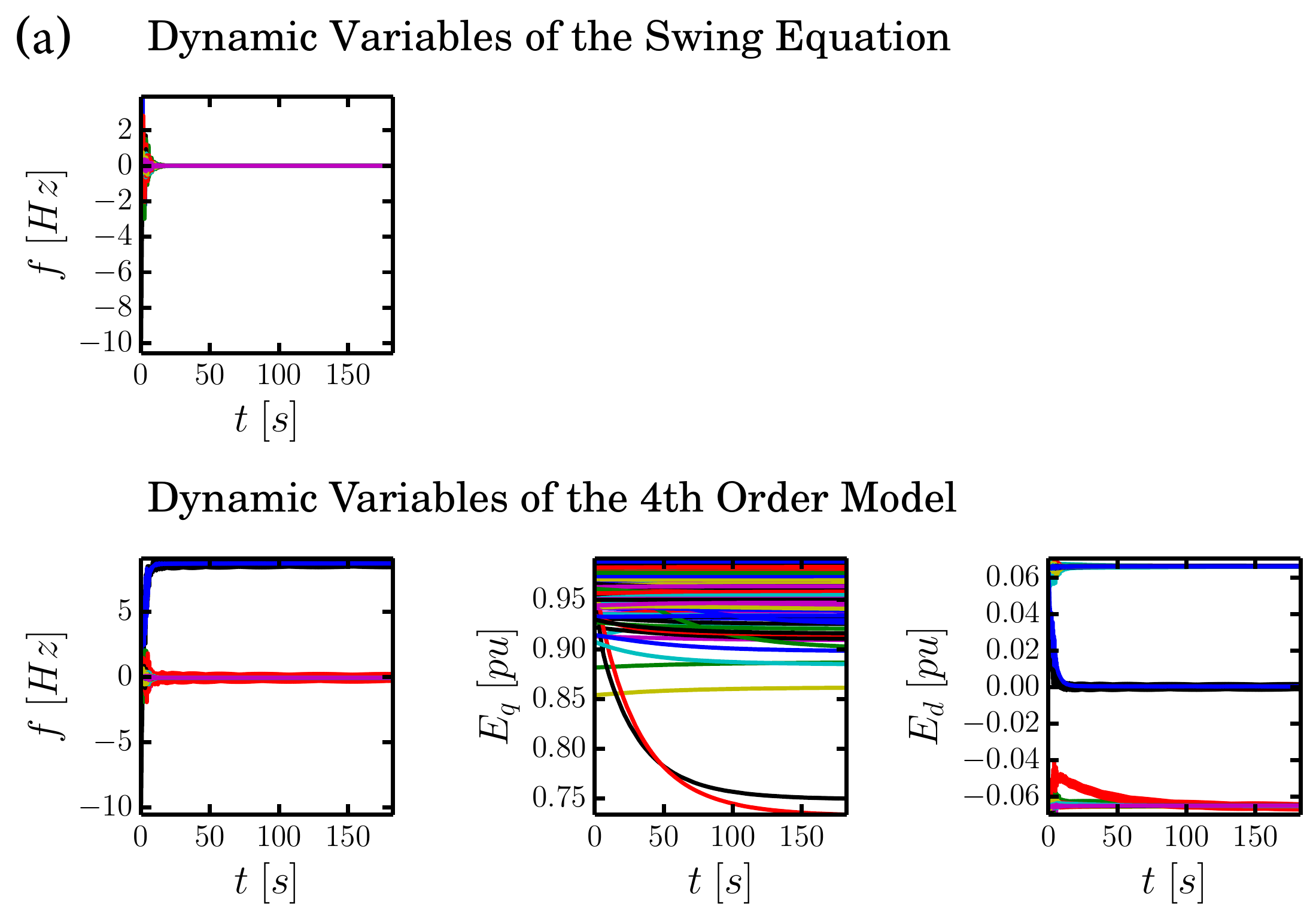}\label{fig:limit_cycle}
\includegraphics[width=0.49\textwidth]{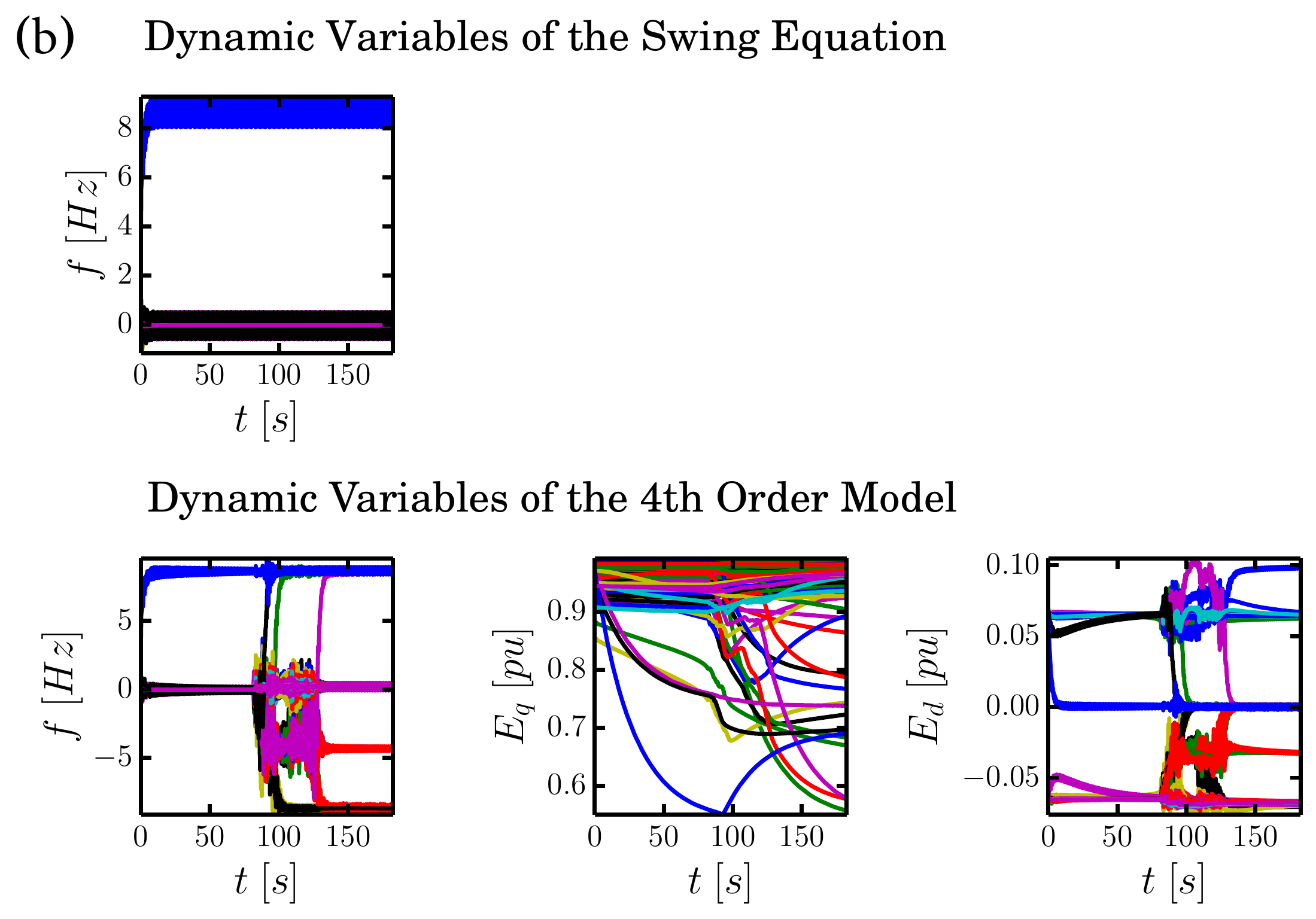}
\caption{(a) Potentially, the systems ends up in different attractors for 2nd order, fix point convergence, and 4th order, limit cycle ($\phi_{104}(0)=1.6$rad, $f_{104}(0)=-10.6$Hz).  (b) Occasionally, the slow voltage dynamics will drive the system into a chaotic transient at around $100$s ($\phi_{231}(0)=2.1$rad, $f_{231}(0)=4.0$Hz). The colors of the trajectories represent the different grid nodes.
\label{fig:time_plots2}}
\end{figure}

Figures \ref{fig:time_plots1} and \ref{fig:time_plots2} contain several example trajectories calculated from single node for the case of the Scandinavian power grid. In many cases, the dynamics behave similarly, and the 4th order model actually converges faster than the swing equation, as far as the frequency is concerned, only while also containing a much slower convergence on the voltage side.

In other cases, such as Fig. \ref{fig:time_plots2}b, the 4th order model seems to be heading towards the same regime as the swing equation for a considerable amount of time, but then enters a new transient regime before settling into a different fixed point. The intermediate time period tends to be around time 400 in most trajectories we observed.

\begin{figure}[tbph!]
\centering
\includegraphics[width=0.9\textwidth]{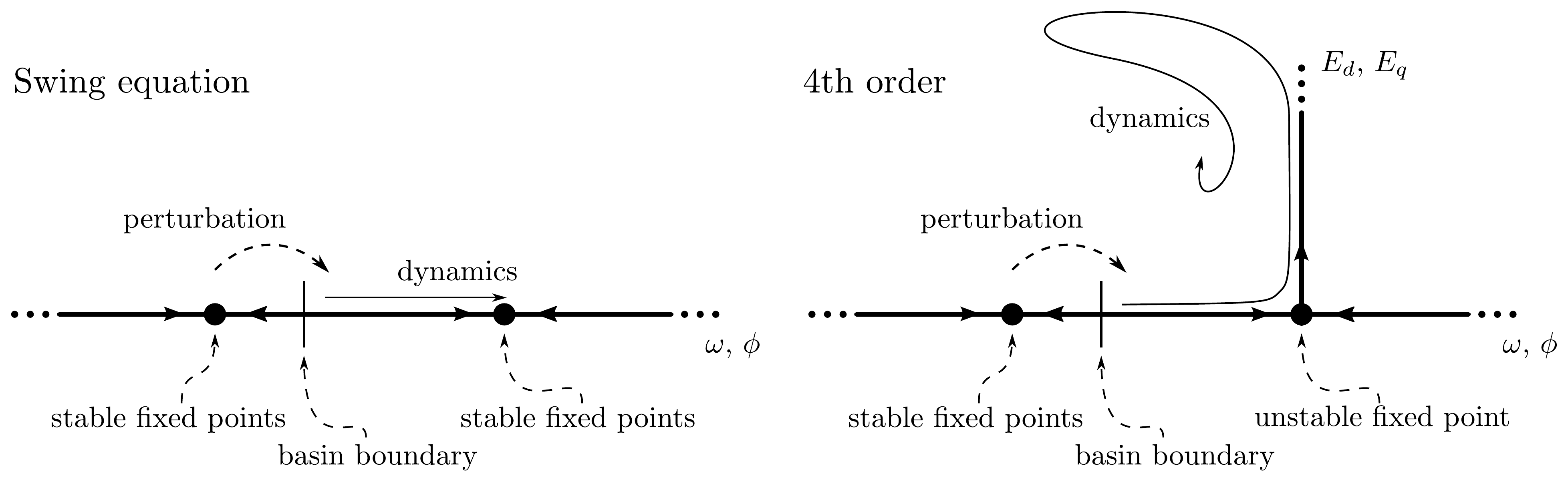}
\caption{A sketch of the dynamics. The fixed point on the right becomes unstable when adding the voltage directions. The perturbation, being purely in the $\phi$, $\omega$ directions, leads us close to the unstable fixed point, but then diverges in the voltage directions until it enters a non-linear regime again. Being close to the unstable fixed point means the dynamics are slow on the transient.
\label{fig:dynamic sketch}}
\end{figure}

This behavior can be understood in terms of unstable fixed points of the 4th order equations. In Fig.  \ref{fig:time_plots2}b, it appears that the trajectory first converges very fast towards a fixed point of the swing equation that is however unstable in the voltage direction. As we have perturbed the system in the dimensions of the swing equation only, we end up getting very close to the unstable fixed point, accounting for a long pseudo convergence, before diverging back into a deeply non-linear regime and finally settling, after a transient of variable length, on a proper attractor (see Fig. \ref{fig:dynamic sketch} for illustration). The transient that leads us back into the non-linear regime is voltage driven and in all cases leads to large voltage deviations. We thus see that there is a connection between large transient voltage deviations and changes in the asymptotic structure.

As noted above, if the Swing equation and the 4th order approach comparable fixed points, the convergence of the 4th order is faster. In the case of limit cycles, the oscillations tend to be smaller. Indeed, the maximal real part of the eigenvalues of the Jacobian of the fourth-order model are always larger than those of the swing equation: $\lambda^{(2)} = -0.1$ whereas we have found that in all systems we studied, $\lambda^{(4)} \in [-0.14,-0.2]$.\par

Generally it is always the case that a limit cycle is associated with a large voltage deviation. Conversely there are cases of large voltage deviations for fixed points, though they are considerably more rare.

\subsection{Basin Stability and Survivability}\label{sec:res_bs_surv}

\begin{figure}[tbph!]
\centering
\includegraphics[width=0.8\textwidth]{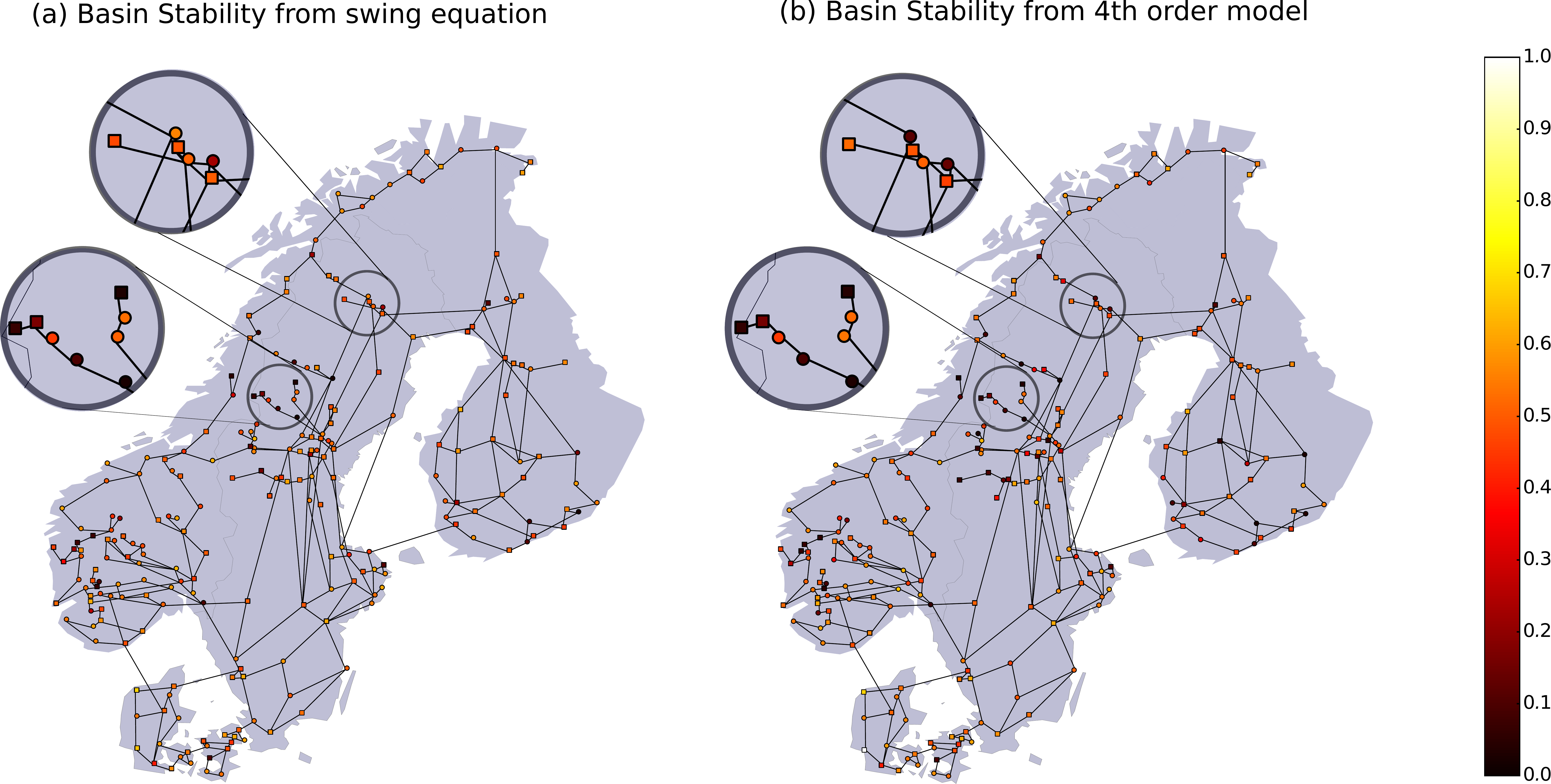}
\caption{Single node basin stability for (a) 2nd and (b) 4th order on the Scandinavian network and frequency disturbances in the range $[-100,100]$. Most nodes do not change much moving from 2nd to 4th order (see inlay I). However, a few (see inlay II and central Finland) become dramatically less stable when including voltage dynamics.}
\label{fig:Scnad_network_deadends}
\end{figure}

In Figure \ref{fig:Scnad_network_deadends} we compare a map for the Scandinavian power grid where each node is colored according to its basin stability value \cite{menck2014dead}. We see that only few individual nodes have dramatically different frequency convergence. Generally, the geographical distribution of basin stability changes only slightly with increasing model detail. 

\begin{figure}[tbph!]
\centering
\includegraphics[width=0.32\textwidth]{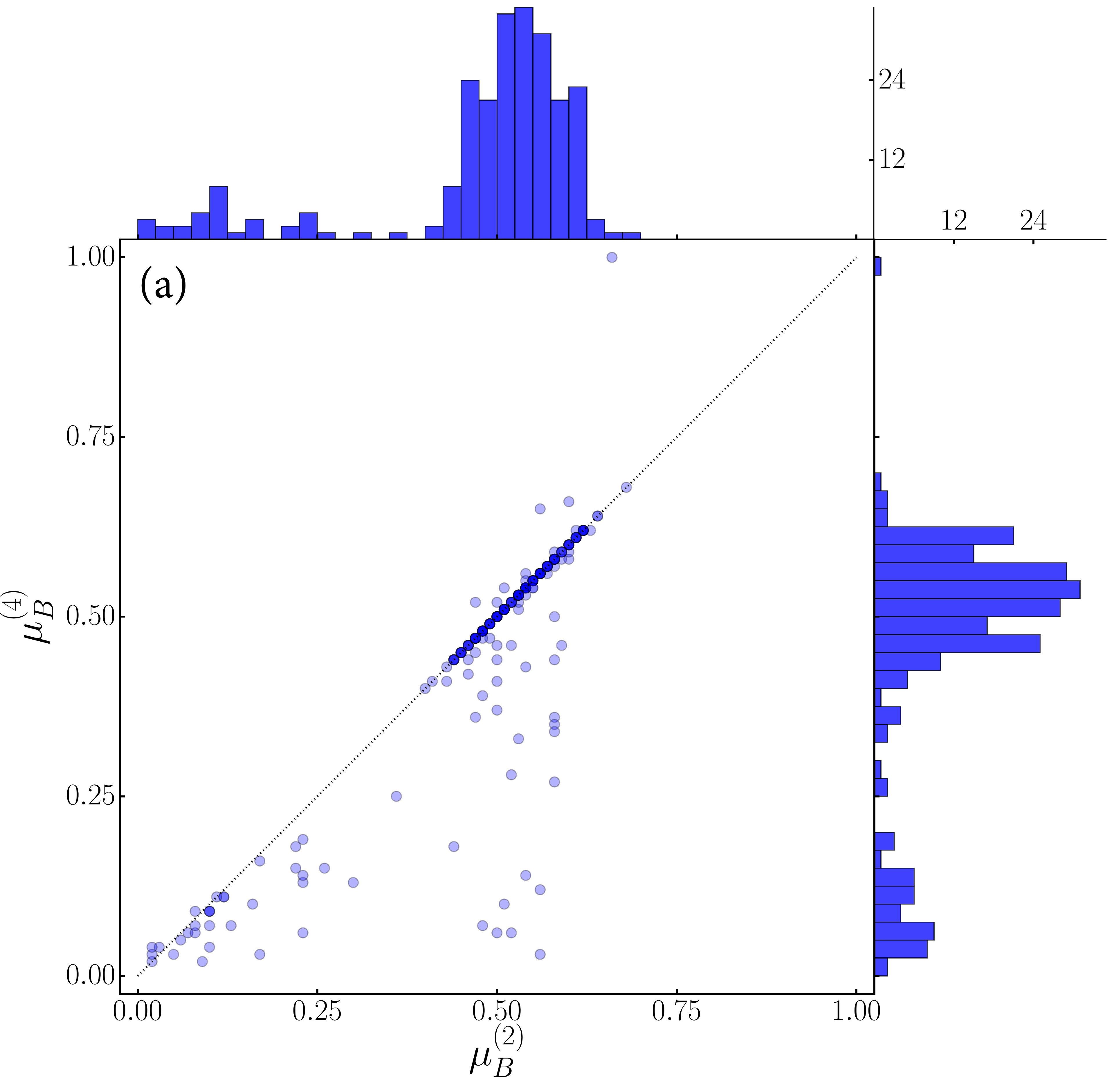}
\includegraphics[width=0.32\textwidth]{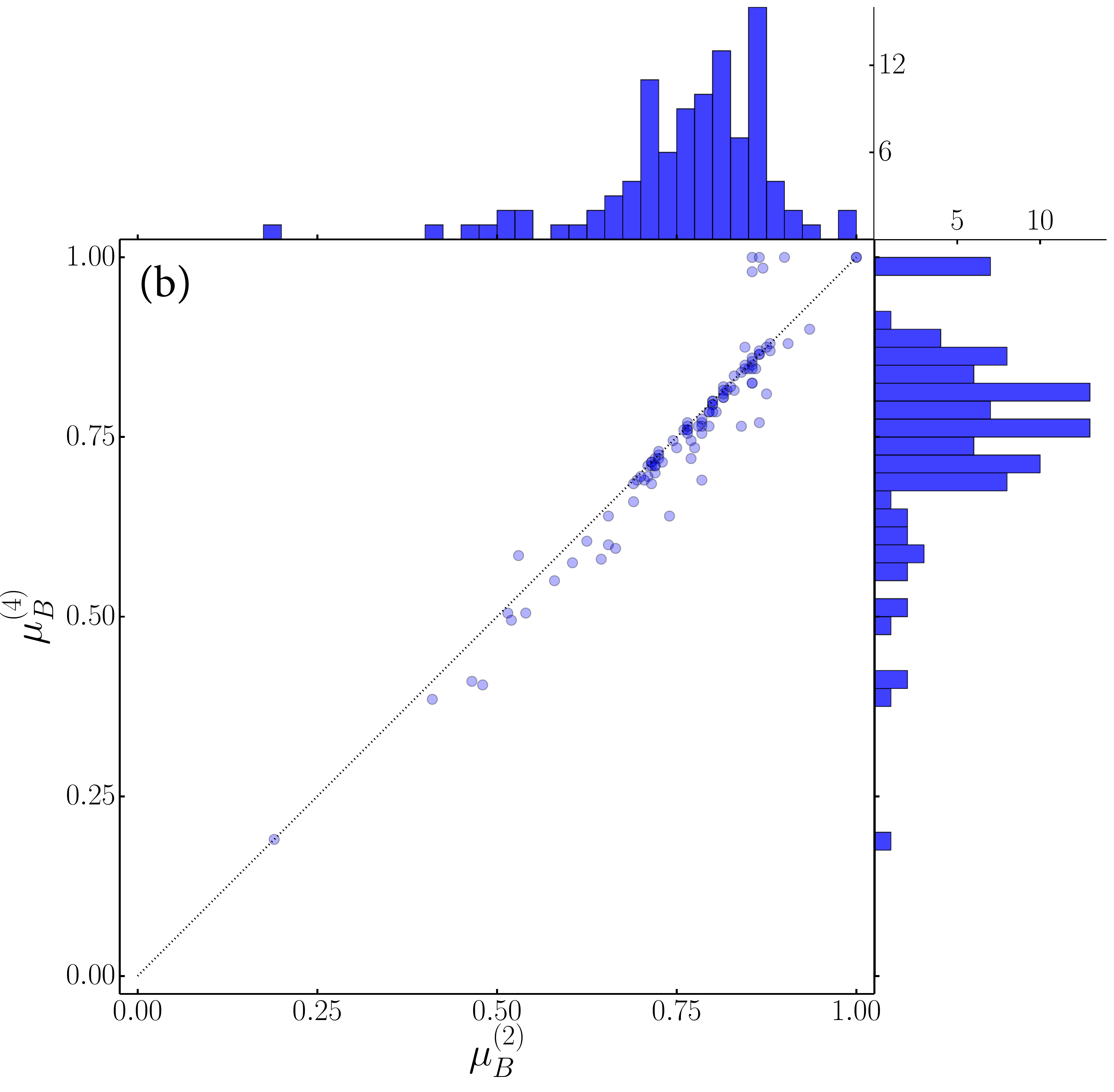}
\includegraphics[width=0.32\textwidth]{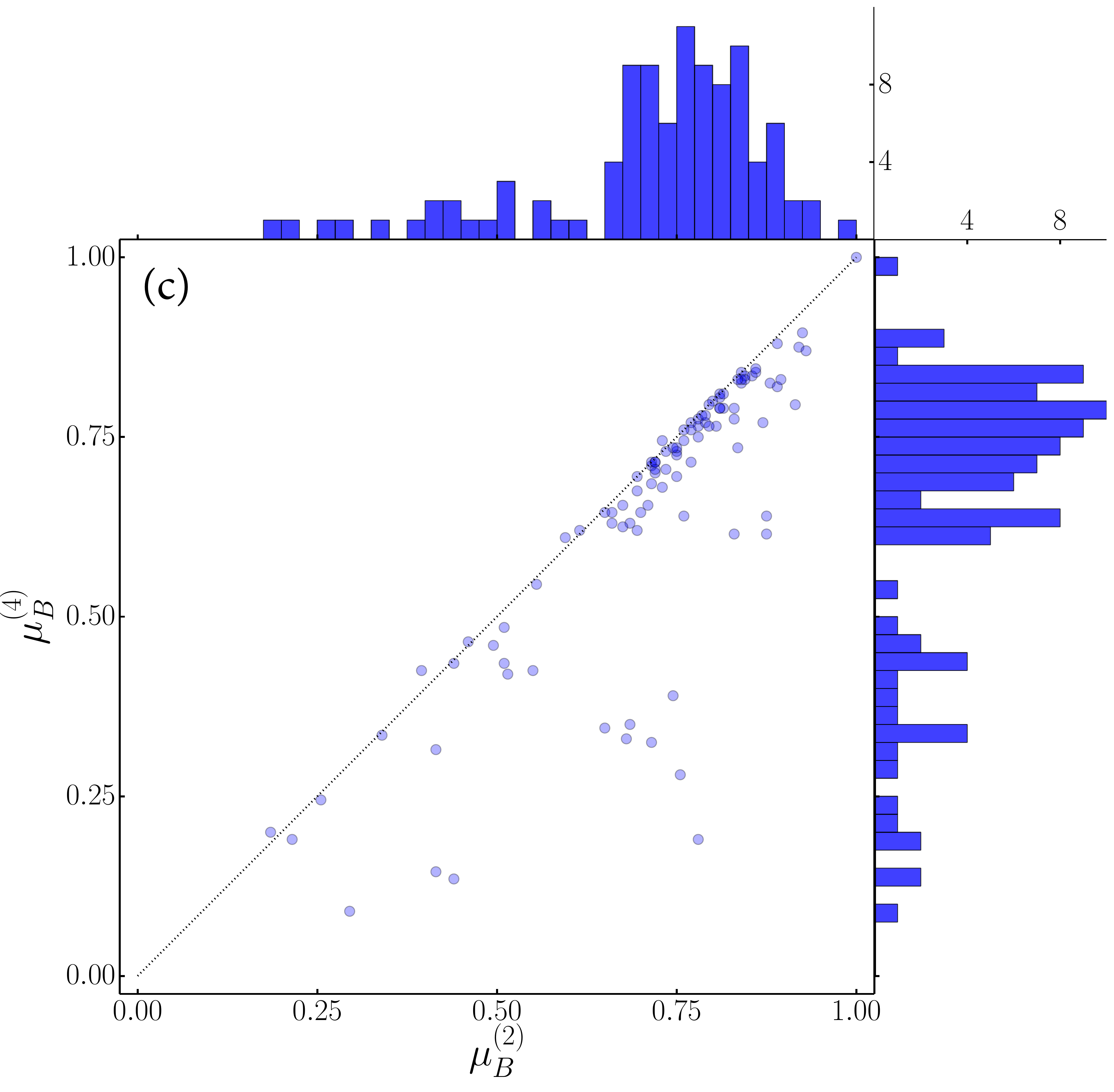}
\caption{Difference in 2nd and 4th order single node basin stability, $\mu^{(2)}_B$ and $\mu^{(4)}_B$, for (a) the Scandinavian grid (maximum frequency perturbation $|f_{max}|=87.2$ Hz), and two synthetic grids ($|f_{max}|=11.0$ Hz): (b) one typical and (c) one extremely divergent. For each grid every node was randomly perturbed $200$ times. \label{fig:scatter basin}}
\end{figure}

In Figure \ref{fig:scatter basin} we show scatter plots of the single node basin stability in the 4th order model versus the Swing equation, for the Scandinavian power grid and two synthetic power grids. We see that the swing equation approximates the stability of power grids well, however, there are some nodes for which it is overestimates. In synthetic grids, with small perturbations, we see that the stability of a small number of already highly stable nodes is boosted by adding voltage dynamics, on the other hand some nodes of average stability drop precipitously when switching to the 4th order. In the Scandinavian power grid, only the latter effect occurs.
We suspect that this is due to a fixed point that is easily reached by a perturbation at that node, becoming unstable. As the comparison of different synthetic networks shows, the existence of such switching fixed points depends heavily on the network structure. Most networks we investigated behave more like the Scandinavian network, with only a few points changing strongly, some however show almost universal deviation.

\begin{figure}[tbph!]
\centering
\includegraphics[width=0.32\textwidth]{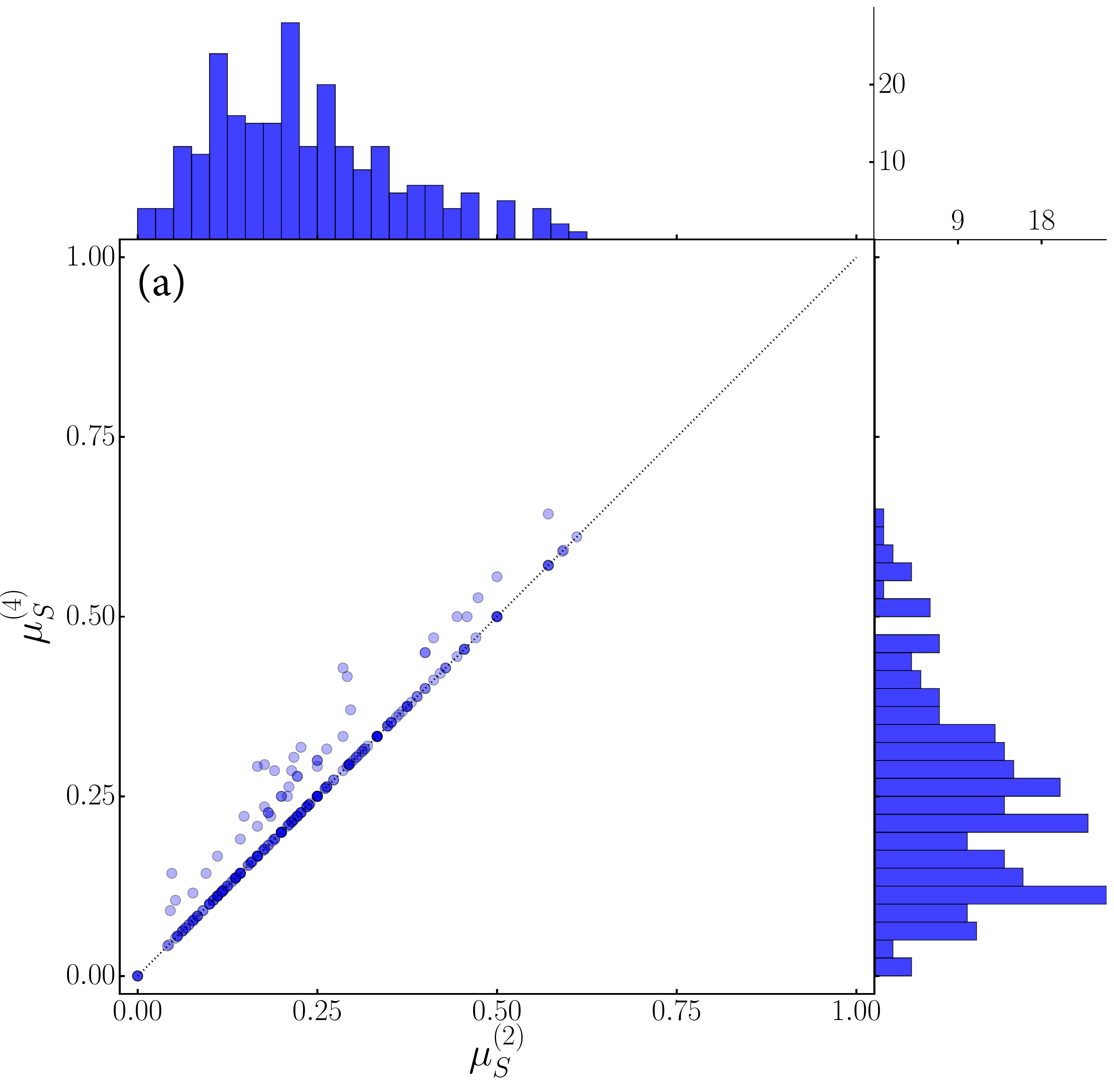}
\includegraphics[width=0.32\textwidth]{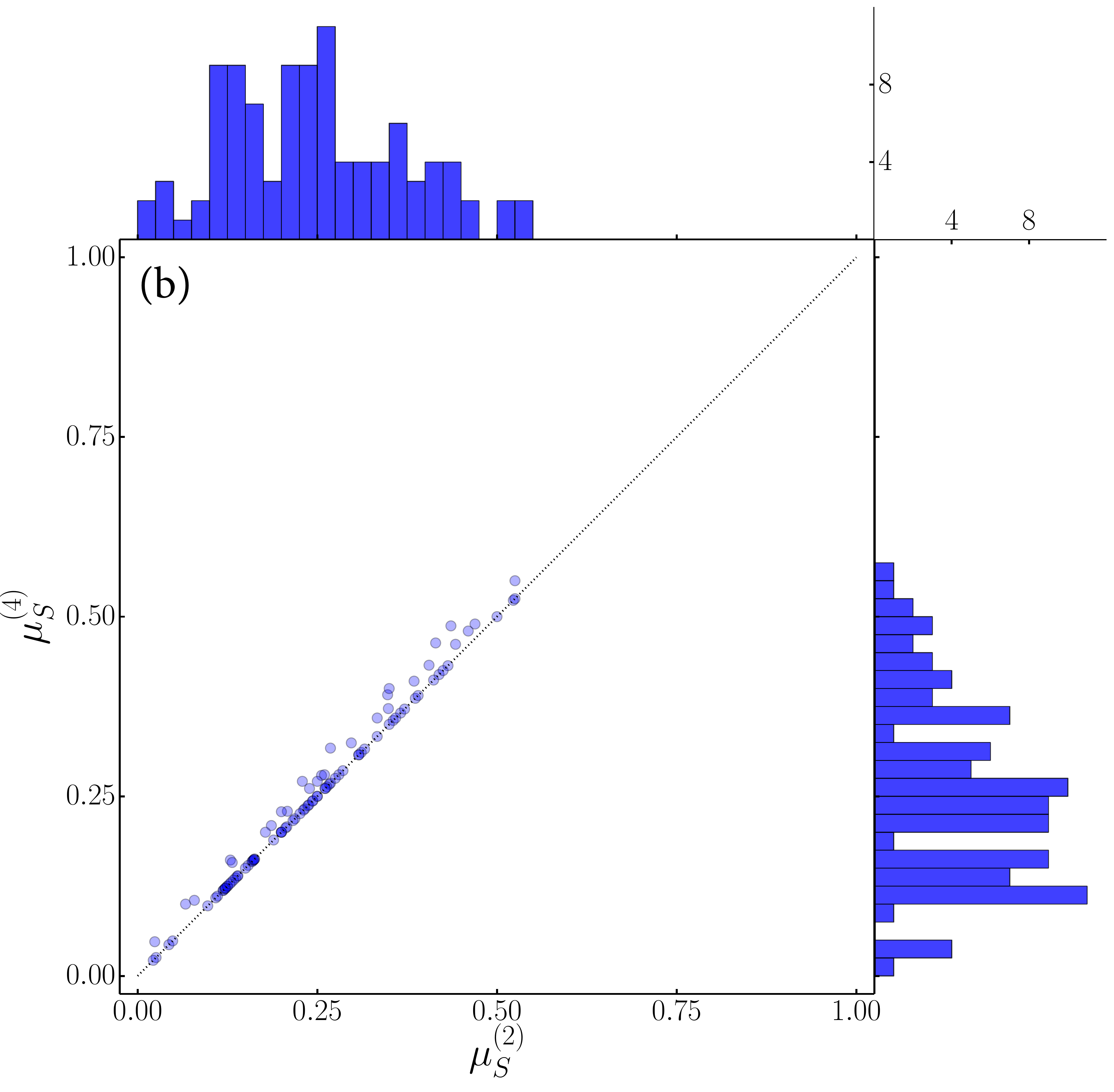}
\includegraphics[width=0.32\textwidth]{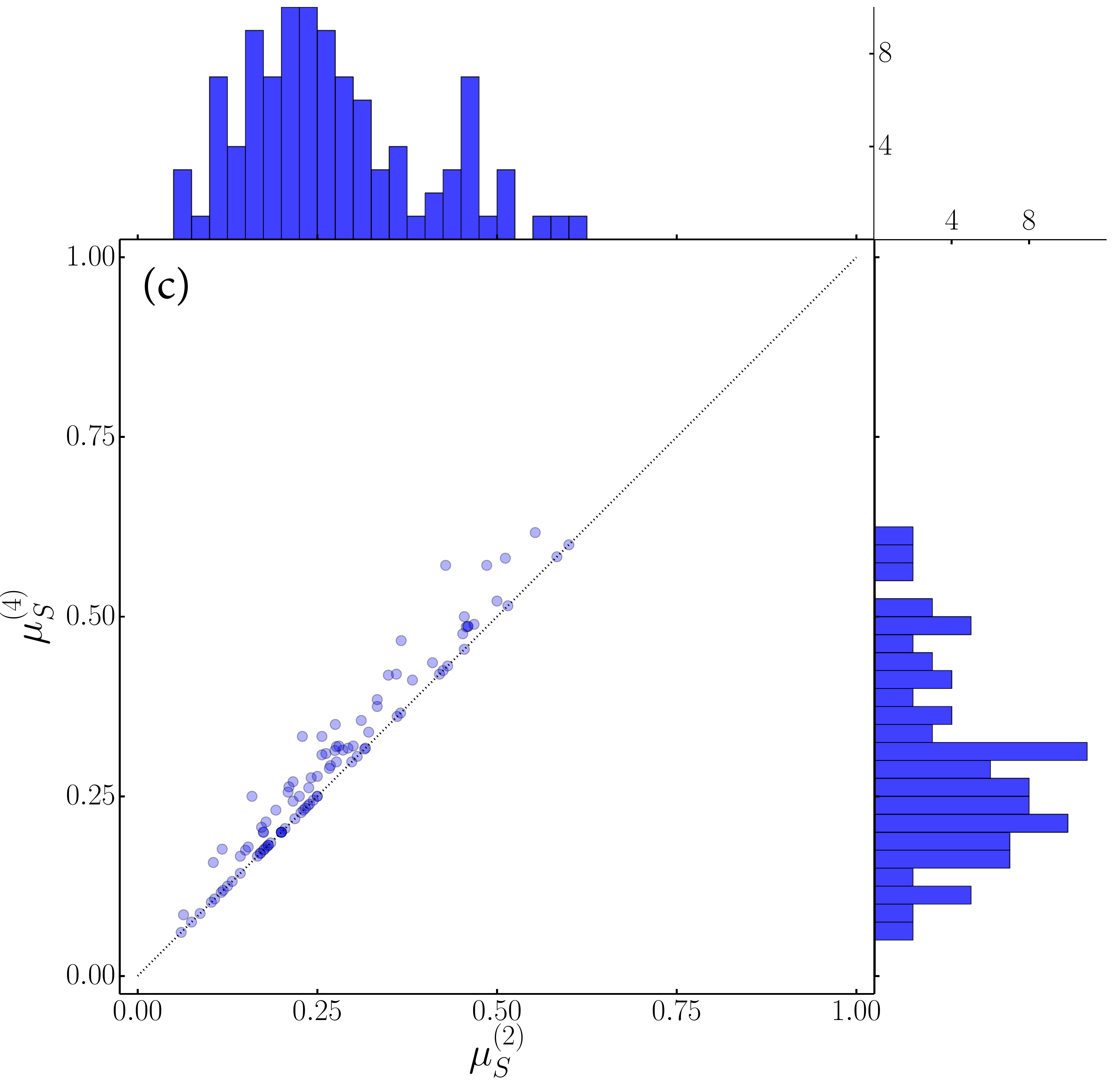}
\caption{Difference in single node survivability, $\mu^{(2)}_S$ and $\mu^{(4)}_S$, for the Scandinavian grid (maximum frequency perturbation $|f_{max}|=87.2$ Hz, critical frequency $|f_{crit}|=8.7$ Hz), and two synthetic grids ($|f_{max}|=11.0$ Hz, $|f_{crit}|=2.2$ Hz), one typical, and one extremely divergent\label{fig:scatter survivability}}
\end{figure}

Conversely, in Fig. \ref{fig:scatter survivability} the survivability for interesting frequency boundaries shows that the voltage dynamic does not affect the maximum deviations during the transient much. The voltage dynamic is apparently to slow to affect the first swing strongly, and the first swing continues to  dominate the transient. Large deviations in the late transients occur mostly when the system is already in a limit cycle, and thus has already violated the frequency bounds.
%

\FloatBarrier

\subsection{Voltage and asymptotic dynamics}\label{sec:volt_bounds}

\begin{figure}[tbph!]
\centering
\includegraphics[width=0.28\textwidth,height= 0.24\textwidth]{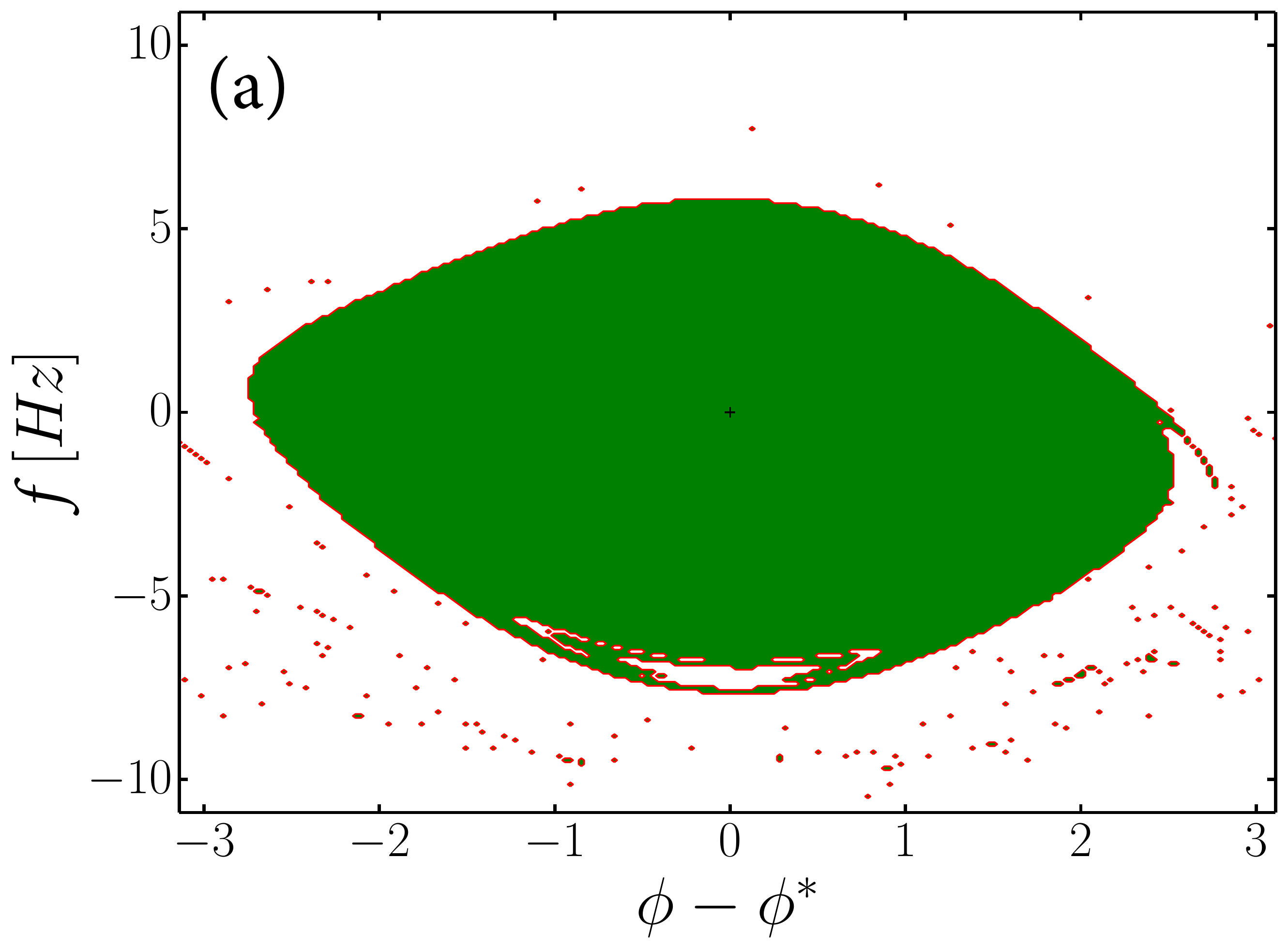}
\hspace{0.05\textwidth}
\includegraphics[height= 0.24\textwidth]{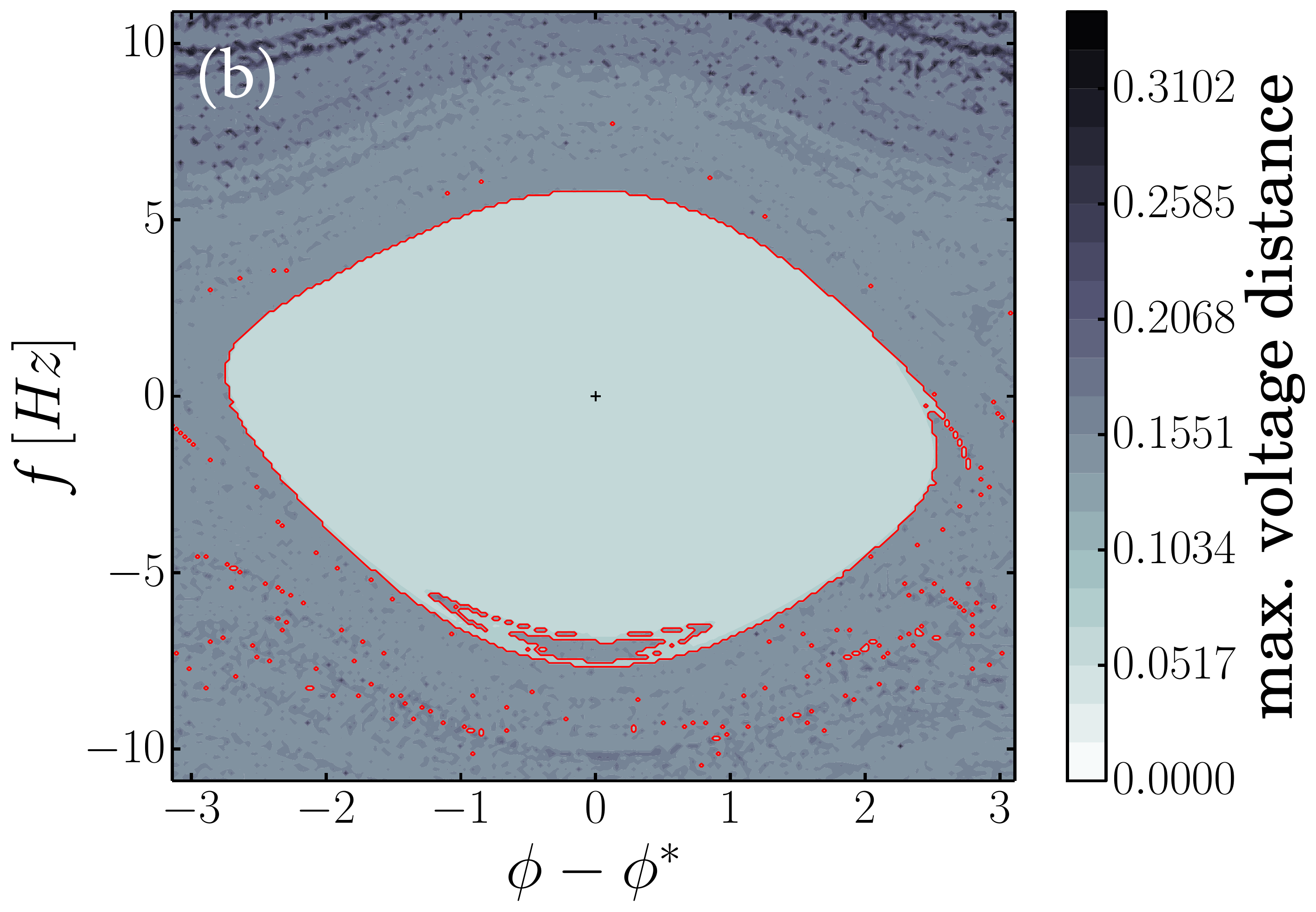}\\
\includegraphics[width=0.35\textwidth]{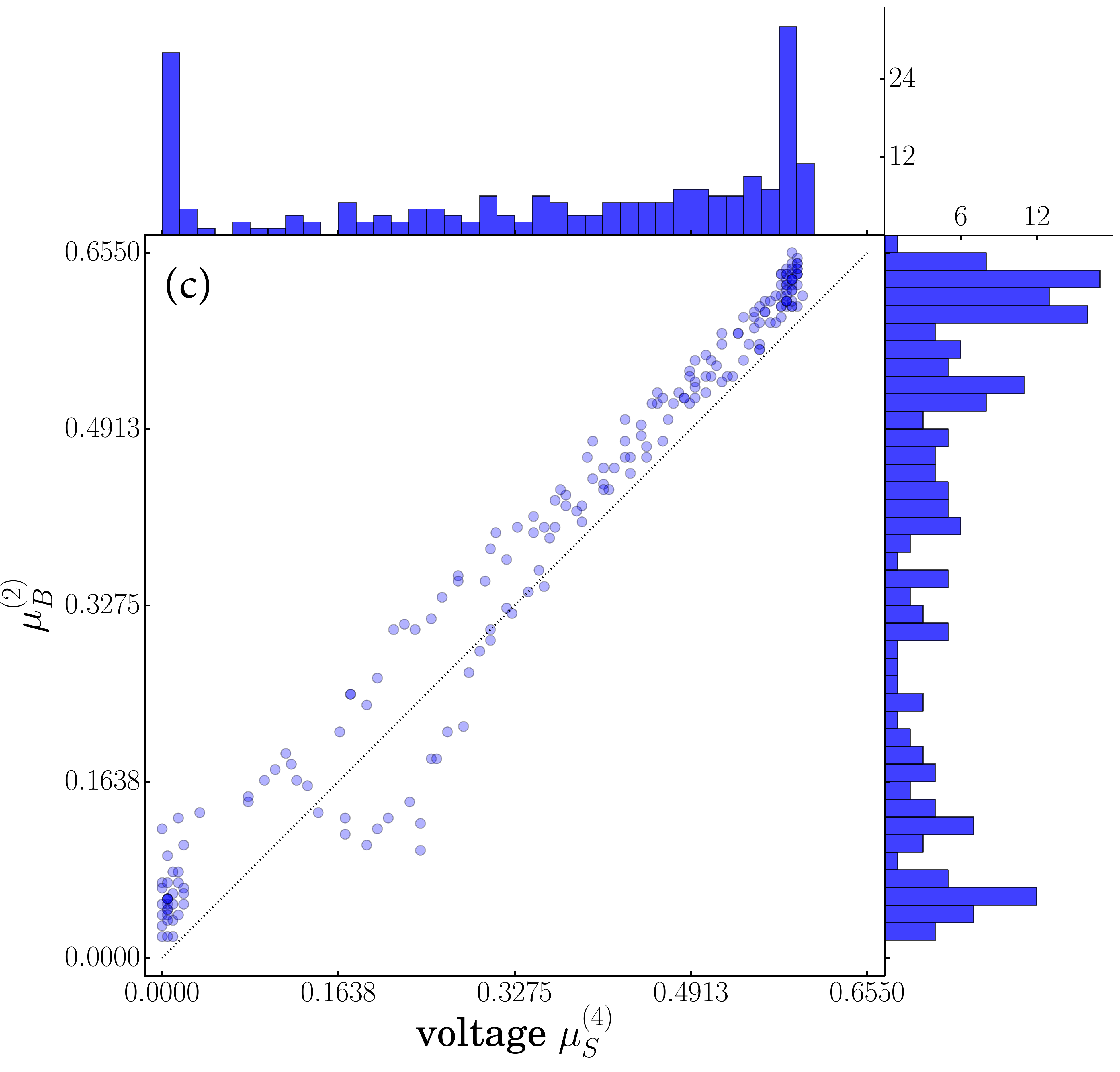}
\includegraphics[width=0.35\textwidth]{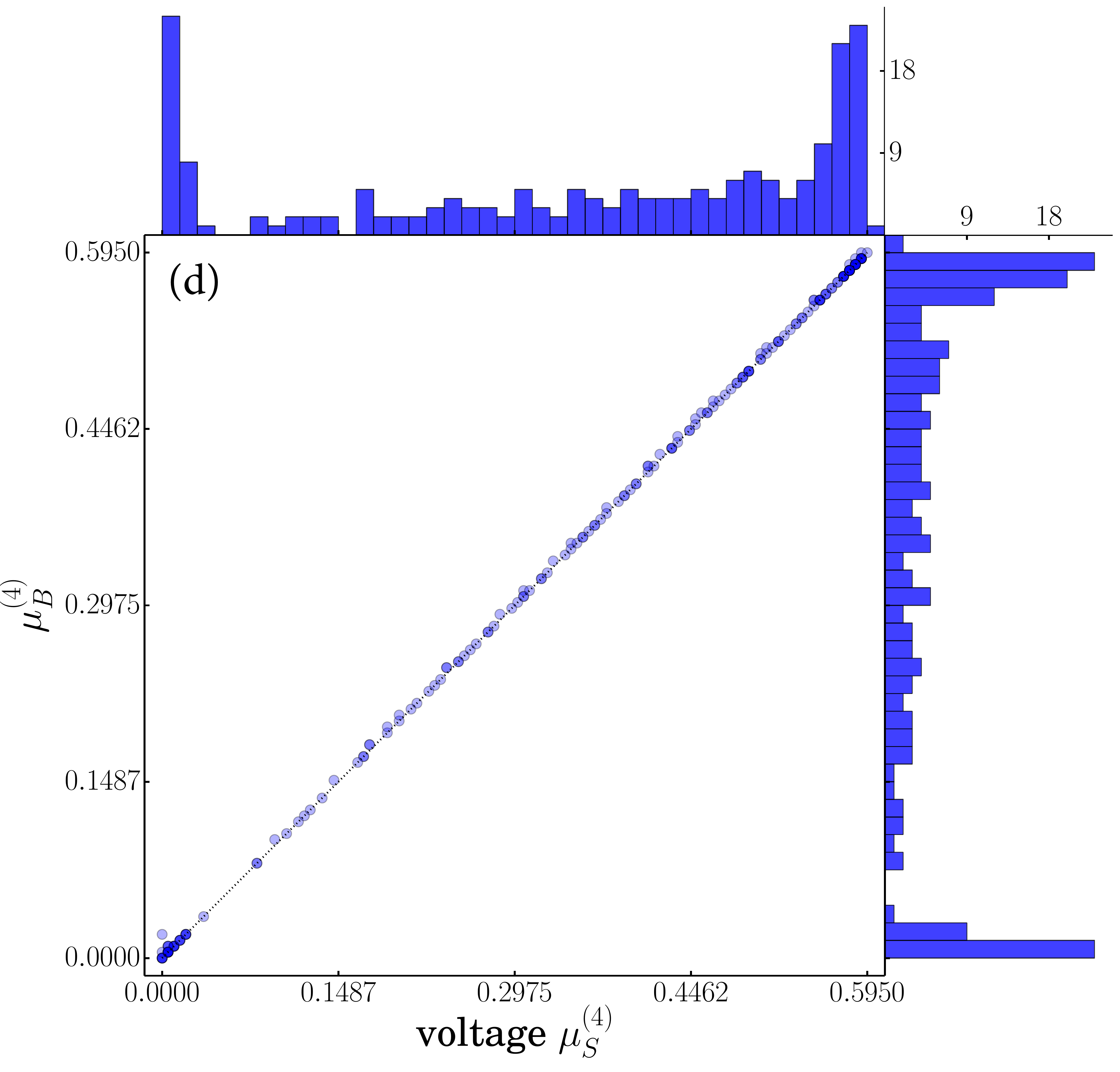}
\caption{(a) Basin of attraction for the fourth order model. (b) Maximum voltage distance for different initial conditions. The red overlay is boundary of the basin of attraction. Comparison of basin of voltage survival of the 4th order model with (c) basin stability for the swing equation and (d) the fourth order model.}
\label{fig:voltageplots}
\end{figure}

We now look at the relationship between voltage transients and asymptotic structures in more detail. The plots in Fig. \ref{fig:voltageplots} are based on single node perturbations in a synthetic power grid. They show that there is an extremely strong relationship between asymptotic behavior and transient voltages. The top left plot shows the basin of attraction (in green) of the fixed points after a perturbation in the frequency and phase at a specific node. The trajectory of top right picture demonstrates how the maximum voltage disturbance follows the perturbation. The red line shows the boundary of the basin of attraction. We see that there is a very distinct step in the maximum transient voltage as we cross the basin boundary. Inside the basin, the maximum voltage disturbance is basically flat, and does not differ noticeably from the one at the fixed point.

After this qualitative reasoning, we also calculated a voltage version of single node survivability which quantifies the ratio of perturbed trajectories for which the voltage disturbance stays below $0.1$pu. In the bottom right plot of Fig.\ref{fig:voltageplots} all nodes of the example synthetic grid have the same ratios of trajectories that stay within voltage bounds and trajectories that show fix point convergence.

This is remarkable as it links an inherently transient, if slow, property with an asymptotic one. Note that as shown in \cite{2015arXiv150601257H} this does not hold for the transient frequency behavior. This also confirms that for the survivability assessment for realistic frequency boundaries, which is entirely concerned with trajectories that return to the fixed point, realistic voltage bounds like $0.1$pu play no role.

\section{Conclusions and outlook}

In this paper we have taken a first step towards applying stability measures from complex systems to more detailed models of the power grid.

In our analysis, we searched for distinct features of the different grid models  (see Section \ref{sec:res_modeldiff}).
Before perturbation, it was necessary to identify the system's stable states first. This search already showed that the fixed point structure of the swing equation and the 4th order model differed more than anticipated. This picture was corroborated by the structure of the late, large transients triggered by slow voltage deviations. These can be understood in schematically as sketched in Fig. \ref{fig:dynamic sketch}. Most remarkably, the frequency and phase perturbation leads to a long trajectory that comes close to a fixed point that is stable in the swing equation but becomes unstable in the 4th order model.

This change in the asymptotic structure implies a large change in the asymptotic behavior whenever such a fixed point becomes relevant. This occured for several nodes in the Scandinavian power grid. Comparisons of several dispatch scenarios for the Scandinavian power grid uncovered a consistent small number of nodes showing these large changes, raising hope that we will be able to identify topological origins of such instabilities in future work.

The survivability in contrast was barely changed between the swing equation and the 4th order model. This can be understood as a consequence of two aspects, first the voltage dynamics is very slow, barely affecting the first swing. The convergence to a fixed point after the first swing are actually faster for the 4th order model than for the swing equation as the largest non-zero eigenvalue is consistently smaller than that of the swing equation by a factor of $1.4$ to $2.0$. The large transients that are triggered late in the system, on the other hand, occur mostly when the system is in a limit cycle regime, and thus has already left the acceptable frequency range.

Maybe the most surprising result was that the transient voltage behavior is intimately linked to the asymptotic structure of the system. Figure \ref{fig:voltageplots} shows the strong connection between voltage transients and asymptotic dynamics in more detail. Astonishingly, the overlay of plots for the basin of attraction and maximum voltage deviation deliver a perfect match (for an example artificial grid). Also, the scatter plot for the survival of voltage for maximum voltage deviations smaller $0.1$pu and basin stability for the 4th order model shows nearly perfect correlation.

Conversely this implies that transient voltage bounds are completely dominated by transient frequency bounds. This further removes the need to take voltage into account for survivability analysis.

\appendix

\section{Parametrization of the Grid Models}\label{sec:app}

In order to facilitate the comparison of our parameter values with the theoretical physics and the engineering literature we give all the reparametrizations done explicitly here:
\vspace{5mm}

%
\begin{tabular}{llll}
\hline\noalign{\smallskip}
Parameter & physical units & reduced units & simulation time units \\
\noalign{\smallskip}\hline\noalign{\smallskip}
$\omega_n, -,\omega^s_n$ & \unit{2\pi\cdot50}{\second^{-1}} &- & \unit{57.357}{\second^{-1}}\\
$H_i $& \unit{3.318}{\second}& - & -\\ 
$T_{d} ,-, T^s_{d}$ & \unit{8.690}{\second} & - &  47.597 \\
$T_{q} ,-, T^s_{q}$& \unit{0.310}{\second} &- & 1.698 \\  
$X_{d} , -, X^s_{d}$& \unit{0.111}{pu} &- & 0.070\\  
$X_{q} , -,X^s_{q}$& \unit{0.103}{pu}& -& 0.065\\  
$D_i , D^r_i , \alpha_i$ & \unit{0.01157}{\second} & \unit{0.548}{\second^{-1}} & 0.1\\ 
$P_i , P^r_i,-$ & \unit{0.6337}{pu}& $\pm$ \unit{30}{\second^{-2}} & $\pm 1$\\ 
$B_{ij} , B^r_{ij} , K_{ij}$ & \unit{0.265}{\ohm\kilo\meter^{-1}} & \unit{0.5}{pu}  & 6\\ 
\hline\noalign{\smallskip}
\end{tabular}

\paragraph{The Swing Equation} is given by

\begin{align}
\frac{d\phi_i}{dt} & = \omega_i \label{eq:swingequation_theta_final_app1}\;,\\
\frac{2H_i}{\omega_{n}}\frac{d\omega_i}{dt} &= P_i - \sum_j U_i B_{ij} U_j \sin(\phi_i - \phi_j) - D_i\omega_i\;.
\end{align}

We work in the per unit (pu) system, and thus the voltage is given in the node dependent unit pu that sets the nominal voltage to $1$. Thus $U_i = 1$pu for the swing equation. The reduced parameters are obtained by absorbing $\frac{2H_i}{\omega_{n}}$ into $D_i$, $B_{ij}$ and $P_i$:

\begin{align}
P_i^{r} &= \frac{P_i \omega_{n}}{ 2 H_i}\nn\\
D_i^{r} &= \frac{D_i \omega_{n}}{ 2 H_i}\nn\\
B_{ij}^{r} &= \frac{B_{ij} \omega_{n}}{ 2 H_i}\nn
\end{align}

These reduced parameters have units $s^{-2}$, $s^{-1}$ and $s^{-2}$pu$^{-2}$ respectively.

For performing simulations, time is often further reparametrized to set $P = \pm1$. We rescale by $\tau: = \beta t$, and thus with $d/dt:=\beta d/d\tau$. This leads to the equations

\begin{align}
\beta \frac{d\phi_i}{d\tau} & = \omega_i \;,\nn\\
\beta^2 \frac{d\omega_i}{d\tau} &= P^r_i - \sum_j U_i B_{ij}^r U_j \sin(\phi_i - \phi_j) - D^r_i\omega_i \nn\;.
\end{align}

With $\omega^s = \frac{\omega}{\beta}$, $P^s_i = \frac{P^r_i}{\beta^2}=\pm1$, $K_{ij} = \frac{B_{ij}^r}{\beta^2}$ and $\alpha_i = \frac{D^r_i}{\beta}$ we obtain

\begin{align}
\frac{d\phi_i}{d\tau} & = \omega^s_i \label{eq:swingequation_theta_app3}\;,\nn\\
\frac{d\omega_i}{d\tau} &= P^s_i - \sum_j U_i K_{ij} U_j\sin(\phi_i - \phi_j) - \alpha_i\omega^s_i \nn,\\
\frac{d\omega_i}{d\tau} &= \pm1 - \sum_j K_{ij} \sin(\phi_i - \phi_j) \unit{1}{pu}^2 - \alpha_i\omega^s_i \nn\;.
\end{align}
%
%

\paragraph{The Fourth-Order Model} was transformed analogously, the voltages are not reparametrized.
\begin{align}
\frac{d\phi_i}{d\tau} & = \omega_i^s\nn,\\
\frac{d^2\phi}{d\tau^2}&= P^s_i - (E_{q,i} I^s_{q,i}-E_{d,i} I^s_{d,i})-\alpha_i\omega_i^s\nn\\
\end{align}

with the current given by
\begin{equation}
I^{s}_i = \sum_{j=1}^{N} K_{ij} U_j e^{i (\phi_j - \phi_i)}.\nn
\end{equation}

The voltage equations are then given in terms of $T^s = \beta T$ and $X^s = \beta^2 \frac{2 H_i}{\omega_{n}} X$:

\begin{align}
T^s_{d,i} \frac{dE_{q,i}}{d\tau} &= -E_{q,i}+X^s_{d,i}I^s_{d,i}+E_f\nn,\\
T^s_{q,i} \frac{dE_{d,i}}{d\tau} &= -E_{d,i}+X^s_{q,i}I^s_{q,i} \nn
\end{align}

Let us summarize all relationships here:

\begin{align}
\beta^2 &= |P_i|&\nn\\
\omega_i &= \omega_i^s \beta&\nn\\
\alpha_i &= \frac{D^r_i}{\beta} = \frac{\omega_n}{2 H_i \beta} D_i\nn\\
K_{ij} &= \frac{B^r_{ij}}{\beta^2} = \frac{\omega_n}{2 H_i \beta^2} B_{ij}\nn\\
P^s_i &= \frac{P^r_i}{\beta^2} = \frac{\omega_n}{2 H_i \beta^2} P_i = \pm1\nn\\
T^s &= \beta T&\nn\\
X^s &= \beta^2 \frac{2 H_i}{\omega_{n}} X&\nn\\
\end{align}

In terms of the reactances found in the literature, the parameter $X$ is defined as the difference between the transient reactance, $X^{'}_{d,q}$, and the static reactance, $X_{d,q}$, in d-/q-axis:
\begin{align}
X_{d,q}  := X_{d,q}-X^{'}_{d,q}
\end{align}
where $X^{'}_{d}$ and $X^{'}_{q}$ are assumed to be equal.
\bibliography{4th_Order_Model_bib}
\bibliographystyle{unsrt}

\end{document}